# Lattice Boltzmann method for relativistic hydrodynamics: Issues on conservation law of particle number and discontinuities

Q. Li,[1] K. H. Luo,[1,2,*] and X. J. Li[3]

[1]Energy Technology Research Group, Faculty of Engineering and the Environment, University of Southampton, Southampton SO17 1BJ, United Kingdom

[2]Center for Combustion Energy, Key Laboratory for Thermal Science and Power Engineering of Ministry of Education, Tsinghua University, Beijing 100084, China

[3]School of Civil Engineering and Mechanics, Xiangtan University, Xiangtan 411105, China

*Corresponding author: K.H.Luo@soton.ac.uk

In this paper, we aim to address several important issues about the recently developed lattice Boltzmann (LB) model for relativistic hydrodynamics [M. Mendoza *et al*., Phys. Rev. Lett. **105**, 014502 (2010); Phys. Rev. D **82**, 105008 (2010)]. First, we study the conservation law of particle number in the relativistic LB model. Through the Chapman-Enskog analysis, it is shown that in the relativistic LB model the conservation equation of particle number is a convection–diffusion equation rather than a continuity equation, which makes the evolution of particle number dependent on the relaxation time. Furthermore, we investigate the origin of the discontinuities appeared in the relativistic problems with high viscosities, which were reported in a recent study [D. Hupp *et al*., Phys. Rev. D **84**, 125015 (2011)]. A multiple-relaxation-time (MRT) relativistic LB model is presented to examine the influences of different relaxation times on the discontinuities. Numerical experiments show the discontinuities can be eliminated by setting the relaxation time $\tau_e$ (related to the bulk viscosity) to be sufficiently smaller than the relaxation time $\tau_v$ (related to the shear viscosity). Meanwhile, it is found that the relaxation time $\tau_\varepsilon$, which has no effect on the conservation equations at the Navier-Stokes level, will affect the numerical accuracy of the relativistic LB model. Moreover, the accuracy of the relativistic LB model for simulating moderately relativistic problems is also investigated.

## I. INTRODUCTION

Relativistic hydrodynamics plays an important role in the fields of astrophysics [1] and high-energy physics [2]. In the literature, many high energy astrophysical phenomena have been investigated by using the relativistic hydrodynamics, such as ultra relativistic jet [3], Neutron star merger [4], and pulsar wind [5]. Moreover, the relativistic hydrodynamics is also of great interest in the context of nuclear physics because of the experiments on heavy-ion (Au-Au) collisions with ultrarelativistic energies at the Relativistic Heavy Ion Collider (RHIC) [6], which reveals a new state of matter: the quark-gluon plasma.

Owing to the high nonlinearity of the relativistic hydrodynamic equations, analytical solutions can be obtained for few simple cases only. Thus in recent years various numerical approaches have been developed for simulating relativistic hydrodynamics. However, the construction of relativistic hydrodynamic equations within the framework of convectional numerical methods usually suffers from several serious problems [7]. For this reason, some numerical formulations based on the kinetic theory have been developed, such as the Boltzmann approach of multiparton scattering (BAMPS) [8], which solves the full Boltzmann equation.

Recently, a relativistic lattice Boltzmann (LB) model was proposed by Mendoza *et al*. [9] for simulating relativistic hydrodynamics. This new model can be treated as a relativistic extension of the standard LB equation [10-13], which is a special discretization scheme of the Boltzmann equation. The relativistic LB model utilizes two different distribution functions, one for the particle number and the other for the energy-momentum. Mendoza *et al.* have validated the relativistic LB model via two relativistic problems, shock waves in quark-gluon plasmas and blast waves from supernova explosions impinging against dense interstellar clouds. In addition, they found that the relativistic LB model is about an order of magnitude faster than the corresponding hydrodynamic codes. Actually, it is well-known that the LB method has many advantages over the conventional numerical methods because of its kinetic origin, the inherent parallelizability on multiple processors, and the avoidance of nonlinear convective terms [14, 15], which makes the relativistic LB model useful in the relativistic context.

Some further studies have also been conducted about the relativistic LB method. Romatschke *et al.* [16] have proposed a fully relativistic LB algorithm, which enables the relativistic LB method capable of dealing with non-Minkowskian geometries and ultrarelativistic fluids. Meanwhile, the extension of the relativistic LB model to the cases with nonideal equation of state has also been made by Romatschke [17]. Recently, through numerical simulations of shock waves in quark-gluon plasma with low and high viscosities, Hupp *et al.* [18] found that the relativistic LB model will lead to unphysical discontinuities in the cases with high viscosities. However, the origin of the discontinuities was not revealed.

In this paper, we aim to further develop the relativistic LB model by addressing several important issues about the model. First, the particle number conservation law in the relativistic LB model will be studied. We will show that the particle number conservation equation in the relativistic LB model is a convection–diffusion equation rather than a continuity equation, which makes the evolution of particle number dependent on the relaxation time. Furthermore, we will investigate the origin of the discontinuities reported in Hupp *et al.*'s study. A multiple-relaxation-time (MRT) relativistic LB model will be presented to examine the influence of different relaxation times. It will be shown that the discontinuities are dependent on the relaxation time $\tau_e$ (related to the bulk viscosity) and can be eliminated by setting $\tau_e$ to be sufficiently smaller than the relaxation time related to the shear viscosity. In addition, the accuracy of the relativistic LB model for simulating moderately relativistic problems will also be investigated.

The rest of the present paper is organized as follows. Section II will briefly introduce the relativistic LB model. The particle number conservation equation of the relativistic LB model will be studied in Sec. III. In Sec. IV, some comments will be made about the energy-momentum conservation equations of the relativistic LB model. In Sec. V, a MRT relativistic LB model will be presented to investigate the origin of the discontinuities appeared in the cases with high viscosities. Finally, a brief conclusion will be made in Sec. VI.

## II. RELATIVISTIC LB MODEL

The relativistic LB model proposed by Mendoza *et al.* is based on the relativistic hydrodynamic equations associated with the conservation of particle number and the energy-momentum conservation. The related energy-momentum tensor is given as follows [9, 19]:

$$T^{\mu\nu} = P\eta^{\mu\nu} + (\varepsilon + P)u^\mu u^\nu + \pi^{\mu\nu}, \tag{1}$$

where $\varepsilon$ is the energy density, $P$ is the hydrostatic pressure, $\eta^{\mu\nu}$ is the Minkowski metric, $\pi^{\mu\nu}$ is the dissipative component of the stress-energy tensor, and $u^\mu$ is the four-vector velocity defined by $u^\mu = (\gamma, \gamma\boldsymbol{\beta})^\mu$, in which $\boldsymbol{\beta} = \boldsymbol{u}/c_l$ is the velocity of the fluid in units of the speed of light and $\gamma = 1/\sqrt{1-|\boldsymbol{u}/c_l|^2}$ is the Lorentz factor. The superscripts $\mu$ and $\nu$ denote the four-dimensional spacetime. They can be identified according to the normal convention when referring to a specific coordinate system, e.g., $\mu = t, x, y, z$ for Cartesian coordinates [20]. The relativistic hydrodynamic equations in Cartesian coordinates can be given by [18, 19]

$$\partial_t (n\gamma) + \partial_i (n\gamma u_i) = 0, \tag{2}$$

$$\partial_t \left((\varepsilon + P)\gamma^2 - P\right) + \partial_i \left((\varepsilon + P)\gamma^2 u_i\right) + \partial_t \pi^{00} + \partial_i \pi^{i0} = 0, \tag{3}$$

$$\partial_t \left((\varepsilon + P)\gamma^2 u_j\right) + \partial_i \left((\varepsilon + P)\gamma^2 u_i u_j\right) + \partial_j P + \partial_t \pi^{0j} + \partial_i \pi^{ij} = 0, \tag{4}$$

where $n$ is the baryon number, which is called particle number in Refs. [9, 18, 19]. The subscripts $i$ and $j$ denote $x$, $y$, and $z$. The index "0" denotes the time component.

To simulate the relativistic hydrodynamic equations, the following two evolution equations with the Bhatnagar-Gross-Krook collision operator are adopted in the relativistic LB model [9, 18, 19]:

$$f_\alpha (\boldsymbol{x} + \boldsymbol{e}_\alpha \delta_t, t + \delta_t) - f_\alpha (\boldsymbol{x}, t) = -\frac{\delta_t}{\tau_f} (f_\alpha - f_\alpha^{eq}), \tag{5}$$

$$g_\alpha (\boldsymbol{x} + \boldsymbol{e}_\alpha \delta_t, t + \delta_t) - g_\alpha (\boldsymbol{x}, t) = -\frac{\delta_t}{\tau_g} (g_\alpha - g_\alpha^{eq}), \tag{6}$$

where $f_\alpha$ is the distribution function for the particle number, $g_\alpha$ is the distribution function for the fluid energy-moment, $\delta_t$ is the time step, $\boldsymbol{e}_\alpha$ are discrete velocities, and $\tau_f$ and $\tau_g$ are the relaxation times for $f_\alpha$ and $g_\alpha$, respectively. The equilibrium distribution functions $f_\alpha^{eq}$ and $g_\alpha^{eq}$ can be determined by the corresponding constraints. For the D3Q19 lattice, $f_\alpha^{eq}$ and $g_\alpha^{eq}$ are given by [18, 19]:

$$f_\alpha^{eq} = w_\alpha n \gamma \left[ 1 + \frac{(\boldsymbol{e}_\alpha \cdot \boldsymbol{u})}{c_s^2} + \frac{(\boldsymbol{e}_\alpha \cdot \boldsymbol{u})^2}{2c_s^4} - \frac{\boldsymbol{u}^2}{2c_s^2} \right], \tag{7}$$

$$g_{\alpha=0}^{eq} = w_\alpha (\varepsilon + P) \gamma^2 \left[ 3 - \frac{P(2+3c_s^2)}{(\varepsilon + P)\gamma^2 c_s^2} - \frac{\boldsymbol{u}^2}{2c_s^2} \right], \tag{8}$$

$$g_{\alpha \geq 1}^{eq} = w_\alpha (\varepsilon + P) \gamma^2 \left[ \frac{P}{(\varepsilon + P)\gamma^2 c_s^2} + \frac{(\boldsymbol{e}_\alpha \cdot \boldsymbol{u})}{c_s^2} + \frac{(\boldsymbol{e}_\alpha \cdot \boldsymbol{u})^2}{2c_s^4} - \frac{\boldsymbol{u}^2}{2c_s^2} \right], \tag{9}$$

where $c_s = c/\sqrt{3}$ ($c$ is the lattice speed). The macroscopic variables are calculated via

$$n\gamma = \sum_\alpha f_\alpha, \quad (\varepsilon + P)\gamma^2 - P = \sum_\alpha g_\alpha, \quad (\varepsilon + P)\gamma^2 \boldsymbol{u} = \sum_\alpha \boldsymbol{e}_\alpha g_\alpha. \tag{10}$$

The shear viscosity is given by $\eta = (\tau_g - 0.5\delta_t)c_s^2(\varepsilon + P)\gamma$.

## III. PARTICLE NUMBER CONSERVATION EQUATION

### A. Theoretical analysis

In this section, we study the particle number conservation equation of the relativistic LB model. In Refs. [9, 19], Mendoza *et al.* claimed that the target conservation equation of particle number, namely Eq. (2), can be recovered from the relativistic LB model. Indeed, there is no doubt that Eq. (2) can be exactly recovered from Eqs. (5) and (7) when the relationships $n\gamma = \sum_\alpha f_\alpha^{eq} = \sum_\alpha f_\alpha$ and $n\gamma \boldsymbol{u} = \sum_\alpha \boldsymbol{e}_\alpha f_\alpha^{eq} = \sum_\alpha \boldsymbol{e}_\alpha f_\alpha$ are satisfied. However, the latter relationship $n\gamma \boldsymbol{u} = \sum_\alpha \boldsymbol{e}_\alpha f_\alpha$ is not satisfied in the relativistic LB model because the velocity in the model is defined by $(\varepsilon + P)\gamma^2 \boldsymbol{u} = \sum_\alpha \boldsymbol{e}_\alpha g_\alpha$.

As a result, in the relativistic LB model the particle number conservation equation will not be a continuity equation. The detailed form can be derived via the Chapman-Enskog analysis, which can be conducted by taking the second-order Taylor series expansion of Eq. (5):

$$\delta_t (\partial_t + \boldsymbol{e}_\alpha \cdot \nabla) f_\alpha + \frac{\delta_t^2}{2}(\partial_t + \boldsymbol{e}_\alpha \cdot \nabla)^2 f_\alpha + O(\delta_t^3) = -\frac{\delta_t}{\tau_f}(f_\alpha - f_\alpha^{eq}), \tag{11}$$

where $\nabla$ is the spatial gradient operator. By introducing the following multiscale expansions

$$\nabla = \kappa \nabla_1, \quad \partial_t = \kappa \partial_{t_1} + \kappa^2 \partial_{t_2}, \quad f_\alpha = f_\alpha^{eq} + \kappa f_\alpha^{(1)} + \kappa^2 f_\alpha^{(2)}, \tag{12}$$

we can rewrite Eq. (11) in the consecutive orders of the expansion parameter $\kappa$ as

$$\kappa: (\partial_{t_1} + \boldsymbol{e}_\alpha \cdot \nabla_1) f_\alpha^{eq} = -\frac{1}{\tau_f} f_\alpha^{(1)}, \tag{13}$$

$$\kappa^2: \partial_{t_2} f_\alpha^{eq} + \left(\partial_{t_1} + \mathbf{e}_\alpha \cdot \nabla_1\right) f_\alpha^{(1)} + \frac{\delta_t}{2}\left(\partial_{t_1} + \mathbf{e}_\alpha \cdot \nabla_1\right)^2 f_\alpha^{eq} = -\frac{1}{\tau_f} f_\alpha^{(2)}. \tag{14}$$

With the aid of Eq. (13), Eq. (14) can be rewritten as

$$\partial_{t_2} f_\alpha^{eq} + \left(1 - \frac{\delta_t}{2\tau_f}\right)\left(\partial_{t_1} + \mathbf{e}_\alpha \cdot \nabla_1\right) f_\alpha^{(1)} = -\frac{1}{\tau_f} f_\alpha^{(2)}. \tag{15}$$

Taking the summations of Eq. (13) and Eq. (15), we can obtain, respectively

$$\partial_{t_1}(n\gamma) + \partial_{1i}(n\gamma u_i) = 0, \tag{16}$$

$$\partial_{t_2}(n\gamma) + \left(1 - \frac{\delta_t}{2\tau_f}\right)\partial_{1j}\left(\sum_\alpha e_{\alpha j} f_\alpha^{(1)}\right) = 0. \tag{17}$$

If the velocity is calculated by $n\gamma\mathbf{u} = \sum_\alpha \mathbf{e}_\alpha f_\alpha$, then $\sum_\alpha e_{\alpha j} f_\alpha^{(1)} = 0$ can be obtained. However, as previously mentioned, such a relationship is not satisfied in the relativistic LB model. According to Eq. (13), Eq. (17) can be rewritten as

$$\partial_{t_2}(n\gamma) = \left(\tau_f - \frac{\delta_t}{2}\right)\partial_{1j}\left[\partial_{t_1}\left(\sum_\alpha e_{\alpha j} f_\alpha^{eq}\right) + \partial_{1i}\left(\sum_\alpha e_{\alpha i} e_{\alpha j} f_\alpha^{eq}\right)\right]. \tag{18}$$

From Eq. (7), we can obtain

$$\partial_{t_1}\left(\sum_\alpha e_{\alpha j} f_\alpha^{eq}\right) = u_j \partial_{t_1}(n\gamma) + n\gamma \partial_{t_1} u_j, \tag{19}$$

$$\partial_{1i}\left(\sum_\alpha e_{\alpha i} e_{\alpha j} f_\alpha^{eq}\right) = u_j \partial_{1i}(n\gamma u_i) + n\gamma u_i \partial_{1i} u_j + c_s^2 \partial_{1j}(n\gamma), \tag{20}$$

where $\partial_{t_1} u_j$ is given by [see Eq. (A27) in the Appendix]

$$\partial_{t_1} u_j = -u_i \partial_{1i} u_j - \partial_{1j} P / \left[(\varepsilon + P)\gamma^2\right]. \tag{21}$$

Substituting Eqs. (16) and (21) into Eq. (19) yields

$$\partial_{t_1}\left(\sum_\alpha e_{\alpha j} f_\alpha^{eq}\right) = -u_j \partial_{1i}(n\gamma u_i) - n\gamma u_i \partial_{1i} u_j - n \partial_{1j} P / \left[(\varepsilon + P)\gamma\right]. \tag{22}$$

According to Eqs. (16) and (18) together with Eqs. (20) and (22), we can obtain

$$\partial_t(n\gamma) + \nabla \cdot (n\gamma\mathbf{u}) = \nabla \cdot \left(\varphi \nabla(n\gamma)\right) - \nabla \cdot (\varphi' \nabla P), \tag{23}$$

where $\varphi = c_s^2(\tau_f - 0.5\delta_t)$ and $\varphi' = nc_s^2(\tau_f - 0.5\delta_t)/\left[(\varepsilon + P)\gamma\right]$.

From Eq. (23) we can see that the particle number conservation equation recovered from the relativistic LB model is a convection–diffusion equation with a source term rather than a continuity equation. In simulations, the diffusion term $\nabla \cdot (\varphi \nabla (n\gamma))$ and the source term $\nabla \cdot (\varphi' \nabla P)$ will result in numerical errors. Since both $\varphi$ and $\varphi'$ are proportional to $(\tau_f - 0.5\delta_t)$, the numerical errors are expected to increase with the increase of $\tau_f / \delta_t$. In addition, it can be seen that the pressure gradient $\nabla P$ will also influence the numerical accuracy.

In previous studies [9, 18, 19], Mendoza *et al*. and Hupp *et al*. used the same relaxation time $\tau$ for $f_\alpha$ and $g_\alpha$, namely $\tau_f = \tau_g = \tau$, and the relaxation time $\tau$ is determined via the shear viscosity $\eta = (\tau - 0.5\delta_t) c_s^2 (\varepsilon + P)\gamma$. Clearly, for relativistic problems with high viscosities, the diffusion and source terms in Eq. (23) will introduce considerable numerical errors. Obviously, to disable these errors, the relaxation time of $f_\alpha$ should be separated from the relaxation time of $g_\alpha$, and must be close to $0.5\delta_t$.

### B. Numerical results

To validate the above analysis, we perform numerical simulations for one-dimensional relativistic shock waves in quark-gluon plasma [9, 18, 19, 21, 22]. For shock waves in viscous quark-gluon matter, the viscosity-entropy density ratio $\eta/s$ is usually used to characterize the problem, and the entropy density $s$ is given by $s = 4n - n \ln \lambda$, where $\lambda = n/n^{eq}$ and $n^{eq} = d_G T^3 / \pi^2$, in which $d_G = 16$ is the degeneration for gluons and $T$ is the temperature [9, 19].

Similar to previous studies, in the present study we also adopt 800 lattices together with open boundaries in the mainstream $z$-direction, and set $c = \delta_x / \delta_t = c_l = 1$. The initial configuration of the simulated problem consists of two regions divided by a membrane at $z = 0$. At $t = 0$, the membrane is removed and the fluid starts expanding. The initial conditions for pressure are given by

$P(z<0) = p_0 = 5.43\,\text{GeV}\,\text{fm}^{-3}$ and $P(z \geq 0) = 2.22\,\text{GeV}\,\text{fm}^{-3}$. In lattice units, the corresponding conditions are $P(z<0) = 2.495 \times 10^{-7}$ and $P(z \geq 0) = 1.023 \times 10^{-7}$. The initial temperature is set to be $T = 350\,\text{MeV}$ (in lattice units $0.0314$) in the whole domain, and the initial particle number is computed with $n = P/T$.

In simulations, three different cases are considered about the relaxation time of $f_\alpha$: $\tau_f = \tau_g$, $\tau_f = 1.0\delta_t$, and $\tau_f = 0.6\delta_t$. The relaxation time $\tau_g$ is determined with $\eta = (\tau_g - 0.5\delta_t)c_s^2(\varepsilon + P)\gamma$. The particle number profiles at $t = 400\delta_t$ (corresponding to $t = 3.2\,\text{fm}/c$) with $\eta/s = 0.01$, $0.05$, and $0.1$ are shown in Fig. 1, which clearly shows that the profiles of the particle number are dependent on the relaxation time $\tau_f$ when the ratio $\eta/s$ is fixed, and it can be seen that, with the increase of $\eta/s$, the differences between the case $\tau_f = \tau_g$ and the other two cases are more and more apparent. Particularly, at $\eta/s = 0.1$, the relaxation time $\tau_g$ is found to be around $15.5\delta_t$, which significantly deviates from $0.5\delta_t$. Consequently, the diffusion term in Eq. (23) will exert an important influence, and this is the reason why the particle number profile of the case $\tau_f = \tau_g$ is much smoother than the profiles of the other two cases. In summary, the theoretical analysis descried in the previous section has been well validated and the numerical results clearly show that the relaxation time of $f_\alpha$ should be close to $0.5\delta_t$ in order to disable the diffusion term.

## IV. ENERGY-MOMENTUM CONSERVATION EQUATIONS

In this section, the energy-momentum conservation equations of the relativistic LB model will be given and some comment will be made. In Ref. [19], Mendoza *et al.* have made a theoretical analysis of the relativistic LB model through the Chapman-Enskog expansion. However, in their analysis some important terms have been omitted. A rigorous Chapman-Enskog analysis of Eq. (6) is therefore provided in the Appendix, which reveals that the relativistic LB model recovers the following energy-momentum conservation equations at the Navier-Stokes level:

$$\partial_t (\sigma - P) + \partial_i (\sigma u_i) = 0, \tag{24}$$

$$\partial_t (\sigma u_j) + \partial_i (\sigma u_i u_j) = -\partial_j P + \partial_i \Pi_{ij} + O(u^3), \tag{25}$$

where $\sigma = (\varepsilon + P)\gamma^2$ and the stress tensor $\Pi_{ij}$ is given by

$$\Pi_{ij} = \eta \left[ \partial_i (\gamma u_j) + \partial_j (\gamma u_i) - \frac{2}{D} \partial_k (\gamma u_k) \delta_{ij} \right] + \varsigma \partial_k (\gamma u_k) \delta_{ij} + (\tau_g - 0.5\delta_t) \chi c_s^2 \partial_k (\sigma u_k) \delta_{ij}$$
$$+ (\tau_g - 0.5\delta_t) \left[ u_j \partial_i (c_s^2 \sigma - P) + u_i \partial_j (c_s^2 \sigma - P) \right] + O(u^3), \tag{26}$$

where $D$ is the spatial dimension, $\eta = (\tau_g - 0.5\delta_t) c_s^2 \sigma / \gamma$ is the shear viscosity, and $\varsigma = 2\eta/D$ is the bulk viscosity, in which $\chi = 1 - P / \left[ c_s^2 (\sigma - P) \right]$.

Now several comments are made about the energy-momentum conservation equations of the relativistic LB model. First, similar to standard LB models, the relativistic LB model has also neglected some third-order velocity terms in deriving the stress tensor Eq. (26) [see Eqs. (A18) and (A24) in the Appendix]. Second, it can be seen that, besides the neglected third-order velocity terms, some other error terms are also included in Eq. (26). These error terms originate from the following changes:

$$\sum_\alpha \hat{g}_\alpha^{eq} = \sigma \Rightarrow \sum_\alpha g_\alpha^{eq} = \sigma - P, \tag{27}$$

$$\sum_\alpha e_{\alpha i} e_{\alpha j} \hat{g}_\alpha^{eq} = \sigma u_i u_j + c_s^2 \sigma \delta_{ij} \Rightarrow \sum_\alpha e_{\alpha i} e_{\alpha j} g_\alpha^{eq} = \sigma u_i u_j + P \delta_{ij}, \tag{28}$$

where $\hat{g}_\alpha^{eq} = w_\alpha \sigma \left( 1 + u_a + 0.5 u_a^2 - 0.5 \mathbf{u}^2 / c_s^2 \right)$, in which $u_a = (\mathbf{e}_\alpha \cdot \mathbf{u}) / c_s^2$.

Obviously, when the inverse changes are made, i.e., when $\sigma - P \Rightarrow \sigma$ and $P = c_s^2 \sigma$, the coefficient $\chi$ will be equal to zero. Then the third and fourth terms on the right-hand side of Eq. (26) will disappear. Actually, the fourth term on the right-hand side of Eq. (26) can be removed by setting $\sum_\alpha e_{\alpha i} e_{\alpha j} e_{\alpha k} g_\alpha^{eq} = P (u_k \delta_{ij} + u_i \delta_{jk} + u_j \delta_{ik})$. However, such a relationship can not be satisfied in the framework of standard lattices (such as D2Q9 and D3Q19) due to their low symmetry [23, 24].

With the regard to the bulk viscosity, it should be noted that in the kinetic theory both the Boltzmann equation and the Boltzmann-BGK equation will give a zero bulk viscosity ($\varsigma = 0$) for monatomic gases. However, the LB-BGK equation, which is a special discretization scheme of the Boltzmann-BGK equation, results in a non-zero bulk viscosity. In the LB community, the non-zero bulk viscosity is usually interpreted as a numerical artifact originating from the influence of the discretization on the attenuation of sound waves [25]. Owing to the nonzero bulk viscosity $\varsigma = 2\eta/D$, for (1 + 1) dimensional relativistic problems ($u_x = u_y = \partial_x = \partial_y = 0$), Eq. (26) will give

$$\Pi_{zz} = 2\eta \partial_z (\gamma u_z) + E_r, \qquad (29)$$

where $E_r$ represent the error terms. From Eq. (29) it can be seen that the coefficient before the term $\partial_z (\gamma u_z)$ is $2\eta$. However, according to the relativistic hydrodynamics, the correct coefficient should be $4\eta/3$ ($\varsigma = 0$ and $D = 3$) [22]. A transformation of the shear viscosity is therefore needed. Such a problem has not been noticed in previous studies.

## V. ORIGIN OF DISCONTINUITIES

### A. MRT relativistic LB model

In this section, the origin of the discontinuities reported in Hupp *et al.*'s study [18] will be investigated with a MRT relativistic LB model. Actually, in the LB community it has been well recognized that the MRT collision operator can overcome some obvious defects of the BGK collision operator, such as fixed Prandtl number and fixed ratio between the shear and bulk viscosities. In addition, much research has shown that the MRT collision operator can improve the numerical stability of LB models by separating the relaxation times of hydrodynamic and non-hydrodynamic moments [26-28, 15, 24].

The MRT-LB equation can be obtained by replacing the BGK collision operator with the MRT collision operator, and then Eq. (6) can be rewritten as

$$g_\alpha(x+e_\alpha\delta_t, t+\delta_t) = g_\alpha(x,t) - \delta_t \Lambda_{\alpha\beta}(g_\beta - g_\beta^{eq})\big|_{(x,t)}, \tag{30}$$

where $\Lambda = M^{-1}SM$ is the collision matrix, $M$ is an orthogonal transformation matrix, and $S$ is a diagonal Matrix. In the present study, we consider (1 + 1) dimensional relativistic problems only. Hence the MRT collision operator based on the D2Q9 lattice is adopted. The corresponding diagonal Matrix $S$ is given by

$$S = \text{diag}(\tau_\sigma^{-1}, \tau_e^{-1}, \tau_\varepsilon^{-1}, \tau_j^{-1}, \tau_q^{-1}, \tau_j^{-1}, \tau_q^{-1}, \tau_v^{-1}, \tau_v^{-1}), \tag{31}$$

where $\tau_\sigma = \tau_j = \delta_t$ are the relaxation times for the conserved moments; $\tau_v$ and $\tau_e$ are the relaxation times related to the shear and bulk viscosities, respectively. For the D2Q9 lattice, according to the related constraints, the equilibrium distribution function $g_\alpha^{eq}$ is defined as follows:

$$g_{\alpha=0}^{eq} = w_\alpha \sigma \left[\frac{9}{4} - \frac{P(5+9c_s^2)}{4\sigma c_s^2} - \frac{u^2}{2c_s^2}\right], \tag{32}$$

$$g_{\alpha\geq 1}^{eq} = w_\alpha \sigma \left[\frac{P}{\sigma c_s^2} + \frac{(e_\alpha \cdot u)}{c_s^2} + \frac{(e_\alpha \cdot u)^2}{2c_s^4} - \frac{u^2}{2c_s^2}\right], \tag{33}$$

where $\sigma = (\varepsilon + P)\gamma^2$. Through the transformation matrix, the distribution function $g_\alpha$ and its equilibrium distribution $g_\alpha^{eq}$ can be projected onto the moment space via $m = Mg$ and $m^{eq} = Mg^{eq}$, respectively, where $g = (g_0, g_1, \cdots, g_8)^T$ and $g^{eq} = (g_0^{eq}, g_1^{eq}, \cdots, g_8^{eq})^T$. After some algebra, the following equilibria $m^{eq}$ can be obtained:

$$m^{eq} = \big[(\sigma - P), (-4\sigma + 3\sigma u^2 + 10P), (4\sigma - 3\sigma u^2 - 13P),$$
$$\sigma u_x, -\sigma u_x, \sigma u_y, -\sigma u_y, \sigma(u_x^2 - u_y^2), \sigma u_x u_y\big]^T, \tag{34}$$

With $m$ and $m^{eq}$, Eq. (30) can be implemented as follows:

$$m^* = m - \delta_t S(m - m^{eq}), \quad g_\alpha(x+e_\alpha\delta_t, t+\delta_t) = g_\alpha^*(x,t), \tag{35}$$

where $g^* = M^{-1}m^*$. The transformation matrix $M$ of the D2Q9 lattice and its inverse matrix $M^{-1}$

can be found in Ref. [24].

The energy-momentum conservation equations recovered form the MRT relativistic LB model can also be obtained with the Chapman-Enskog analysis, which can be implemented in the moment space [29]. The recovered conservation equations are the same as those given in Eqs. (24) and (25) expect that the stress tensor is given by

$$\Pi_{ij} = \eta \left[ \partial_i (\gamma u_j) + \partial_j (\gamma u_i) - \frac{2}{D} \partial_k (\gamma u_k) \delta_{ij} \right] + \varsigma \partial_k (\gamma u_k) \delta_{ij} + \chi (\tau_e - 0.5\delta_t) c_s^2 \partial_k (\sigma u_k) \delta_{ij}$$
$$+ (\tau_v - 0.5\delta_t) \left[ u_j \partial_i (c_s^2 \sigma - P) + u_i \partial_j (c_s^2 \sigma - P) \right] + O(u^3). \quad (36)$$

Here the shear viscosity $\eta = (\tau_v - 0.5\delta_t) c_s^2 \sigma / \gamma$, the bulk viscosity $\varsigma = (\tau_e - 0.5\delta_t) c_s^2 \sigma / \gamma$, $D = 2$, and $c_s^2 = 1/3$. Note that the third term on the right hand side of Eq. (36), which can be rewritten as $\chi \varsigma \gamma \sigma^{-1} \partial_k (\sigma u_k) \delta_{ij}$, is also related to the bulk viscosity. In addition, due to the spatial limit of the 2D MRT collision operator ($D = 2$), a similar transformation of the shear viscosity is also needed when comparing the present numerical results with the results of BAMPS.

### B. Numerical results

Now numerical simulations are carried out for relativistic shock waves in quark-gluon plasma to investigate the discontinuities caused by the relativistic LB model. The configuration of the problem and the initial conditions are the same as those in Sec. III. In simulations, a grid size of $N_x \times N_z = 4 \times 800$ is adopted. The open boundaries are employed in the $z$-direction and the periodic conditions are applied in the $x$-direction. For the MRT relativistic LB model, in addition to $\tau_v$, there are three adjustable relaxation times: $\tau_q$, $\tau_e$, and $\tau_\varepsilon$. To examine the influences of different relaxation times, we consider the following four cases: Case A: $\tau_e = \tau_\varepsilon = \tau_q = \tau_v$; Case B: $\tau_e = \tau_\varepsilon = \tau_v$, $\tau_q - 0.5\delta_t = a(\tau_v - 0.5\delta_t)$; Case C: $\tau_\varepsilon = \tau_q = \tau_v$, $\tau_e - 0.5\delta_t = a(\tau_v - 0.5\delta_t)$; and Case D:

$\tau_q = \tau_v$, $\tau_e = \tau_\varepsilon = 0.5\delta_t + a(\tau_v - 0.5\delta_t)$. The parameter $a$ is chosen as $1/20$. Its influences will be shown later.

For the present test, the corresponding $\beta = u_z/c_l$ is around 0.2, and is in the weakly relativistic regime ($1 < \gamma \ll 2$), in which the third-order velocity terms can be neglected. The predicted velocity profiles of Cases A and B at $t = 400\delta_t$ (i.e., $t = 3.2\,\text{fm}/c$) with $\eta/s = 0.5$ are shown in Figs. 2(a) and 2(b), respectively. For comparison, the results of BAMPS are also presented. From the figure it can be found that a discontinuity at $z = 0$ appears in the both cases. Moreover, it is seen that there are no obvious differences between the results of the two cases. In the LB-MRT community, it is known that the relaxation time $\tau_q$ will affect the numerical accuracy of LB models via the treatment of non-slip boundaries [30]. Nevertheless, the present problem contains no non-slip boundaries and this may be the reason why $\tau_q$ has no effect on the present problem.

The velocity profile of Case C is displayed in Fig. 2(c), from which we can see that the discontinuity appeared in Cases A and B has disappeared. According to the setup of Case C, it can be found that the discontinuity is dependent on the relaxation time $\tau_e$. In other words, the discontinuity is related to the moment $m_1$, which, at the Navier-Stokes level, leads to the second and third terms on the right-hand side of Eq. (36). Clearly, these terms will affect the performance of the relativistic LB model. More importantly, the higher-order terms (beyond the Navier-Stokes level) resulting from the moment $m_1$ will also impose a significant influence, which makes the relaxation time $\tau_e$ pretty important in simulations. For problems with high viscosities, when $\tau_e = \tau_v$ ($\tau_v/\delta_t$ is around 100 for the present problem), the error terms arising from the moment $m_1$ will be of the same order of magnitude as the shear viscosity terms. To damp these errors, the relaxation time $\tau_e$ should be sufficiently smaller than $\tau_v$, and then the unphysical discontinuity can be eliminated.

The velocity profile of Case D is illustrated in Fig. 2(d). By comparing Fig. 2(d) with Fig. 2(c), we can see that the results of Case C deviate from the results of BAMPS in several regions, while the results of Case D basically agree well with those of BAMPS in the whole domain, which indicates that the relaxation time $\tau_\varepsilon$ affects the numerical accuracy of the relativistic LB model. In the literature, it has been shown that $\tau_\varepsilon$ will affect the numerical stability of LB models [30]. Meanwhile, through the Chapman-Enskog analysis it can be easily found that $\tau_\varepsilon$ does not affect the macroscopic equations at the Navier-Stokes level [29], which implies that the influences of $\tau_\varepsilon$ are attributed to the higher-order terms given by the moment $m_2$.

The effects of the parameter $a$ are displayed in Fig. 3 by taking Case D as an example. From the figure we can find that the discontinuity appeared at $z=0$ can be gradually removed with the decrease of the parameter $a$, and it can be seen that there are only several minor differences between the results of $a=1/20$ and $a=1/50$, which means that $a=1/20$ is sufficient for the present problem. The predicted pressure distributions at $\eta/s=0.2$ and $\eta/s=0.5$ are descried in Figs. 4 and 5, respectively. The results of BAMPS given in Ref. [21] are also presented for comparison. Similar discontinuities can be observed in the pressure profiles of Cases A and B, and it can be found that the discontinuities will become strong with the increase of $\eta/s$. Furthermore, good agreement with the results of BAMPS can also be observed in Case D.

Finally, a moderately relativistic shock wave in quark-gluon plasma is also considered. The initial pressure distribution is given by $P(z<0)=5.43\,\text{GeV}\,\text{fm}^{-3}$ and $P(z\geq 0)=0.339\,\text{GeV}\,\text{fm}^{-3}$, which correspond to $2.495\times 10^{-7}$ and $1.557\times 10^{-8}$ in lattice units, respectively [18]. The initial temperature is $T(z<0)=400\,\text{Mev}$ and $T(z\geq 0)=200\,\text{Mev}$ (in lattice units 0.018). The velocity and pressure profiles of Cases A and D at $\eta/s=0.5$ and $t=400\delta_t$ (i.e., $t=3.2\,\text{fm}/c$) are depicted in Fig. 6. In this test the parameter $a$ is set to be $1/30$. From the figure we can see that the results of Case D still obviously deviate from the results of BAMPS although the discontinuities at $z=0$ have been eliminated.

In fact, from Fig. 6 it can be found that the main deviations are located in the region $z/L = [0, 0.375]$, which corresponds to $\beta = u_z/c_l \sim 0.55$ ($c_l = 1$). It is therefore believed that the error terms of the relativistic LB model have imposed an important influence on the numerical results. To identify the influence of the error terms, we introduce a correction term into Eq. (30), and then the collision process can be rewritten as [24]:

$$\mathbf{m}^* = \mathbf{m} - \delta_t \mathbf{S}(\mathbf{m} - \mathbf{m}^{eq}) + \left(\mathbf{I} - \frac{\delta_t \mathbf{S}}{2}\right)\mathbf{C}, \tag{37}$$

where $\mathbf{C} = (C_0, C_1, \cdots, C_8)^{\mathbf{T}}$ is the correction term in the moment space. Strictly speaking, in order to remove the error terms, both the moments $m_1$ and $m_7$ should be corrected for the present problem. However, since the errors resulting from the moment $m_1$ can be damped via $\tau_e$, we can consider the correction of $m_7$ only ($C_{0,1,\cdots 6,8} = 0$). According to the Chapman-Enskog analysis, $C_7$ is given by

$$C_7 = 2\delta_t \left[ u_z \partial_z \left(c_s^2 \sigma - P\right) - c_s^2 \sigma \gamma^{-1} u_z \partial_z \gamma - u_z \partial_z \left(\sigma u_z^2\right) + u_z^2 \vartheta \partial_z \left(\sigma u_z\right) \right], \tag{38}$$

where $\vartheta = 1 + \left[(\varepsilon/P + 1)\gamma^2 - 1\right]^{-1}$. The first term on the right-hand side of Eq. (38) is caused by the change in Eq. (28), the second term is due to the change from $\gamma \partial_z u_z$ to $\partial_z (\gamma u_z)$, while the last two terms are the third-order velocity terms resulting from $\partial_{t_1} (\sigma u_z^2)$. Here it should be noted that Eqs. (37) and (38) mainly serve as a strategy to examine the influence of the error terms. For practical applications, a more sophistical model is required because there will be many error terms in three-dimensional problems. The corrected numerical results of Case D are shown in Fig. 7, from which an obvious improvement can be observed although there are still some minor differences between the corrected results and the results of BAMPS. Actually, these differences are acceptable on the basis of the fact that the LB equation is a special approximation of the full Boltzmann equation.

## VI. CONCLUSION

In this paper, several important issues about the recently proposed relativistic LB model have been theoretically and numerically studied. Firstly, we have shown that the particle number conservation equation of the relativistic LB is a convection–diffusion equation rather than a continuity equation claimed in previous studies. To disable the related error terms, the relaxation time of $f_\alpha$ should be close to $0.5\delta_t$. Secondly, the origin of the discontinuities reported in Ref. [18] has been investigated by using a MRT relativistic LB model. It is found that the discontinuities are dependent on the relaxation time $\tau_e$, and the relaxation time $\tau_\varepsilon$ is found to affect the numerical accuracy of the relativistic LB model although it has no effect on the conservation equations at the Navier-Stokes level. In particular, numerical experiments show that, by setting $\tau_e$ and $\tau_\varepsilon$ to be sufficiently smaller than the relaxation time $\tau_\nu$, the discontinuities appeared in the relativistic problems with high viscosities can be eliminated and the accuracy of the relativistic LB model can be improved.

Furthermore, we have shown that the relativistic LB model will lead to considerable numerical errors for moderately relativistic problems although the discontinuities can be eliminated with the MRT collision operator. Nevertheless, it is also found that the accuracy of the relativistic LB model can be obviously improved when the error terms are removed. In fact, most of the error terms can be removed via $\sum_\alpha e_{\alpha i} e_{\alpha j} e_{\alpha k} g_\alpha^{eq} = \sigma u_i u_j u_k + P\left(u_k \delta_{ij} + u_i \delta_{jk} + u_j \delta_{ik}\right)$. However, this relationship can not be satisfied in the framework of standard lattices and high-order lattices must be used. In other words, a high-order MRT LB model is needed for simulating moderately relativistic problems. In the literature, there have been several high-order MRT LB models for non-relativistic thermodynamics [31, 32], which may offer some insights about constructing high-order MRT LB models for relativistic hydrodynamics. This issue can be considered in future studies.

**ACKNOWLEDGMENTS**

Support by the Engineering and Physical Sciences Research Council of the United Kingdom under Grant No. EP/I012605/1 is gratefully acknowledged.

**APPENDIX: CHAPMAN-ENSKOG ANALYSIS OF EQ. (6)**

In this appendix, a rigorous Chapman-Enskog analysis of Eq. (6) is provided. According to the second-order Taylor series expansion, the following equation can be obtained from Eq. (6):

$$\delta_t \left( \partial_t + \boldsymbol{e}_\alpha \cdot \nabla \right) g_\alpha + \frac{\delta_t^2}{2} \left( \partial_t + \boldsymbol{e}_\alpha \cdot \nabla \right)^2 g_\alpha = -\frac{\delta_t}{\tau_g} \left( g_\alpha - g_\alpha^{eq} \right) + O\left( \delta_t^3 \right). \tag{A1}$$

Using Eq. (12) and the expansion $g_\alpha = g_\alpha^{eq} + \kappa g_\alpha^{(1)} + \kappa^2 g_\alpha^{(2)}$, we can obtain

$$\kappa: \left( \partial_{t_1} + \boldsymbol{e}_\alpha \cdot \nabla_1 \right) g_\alpha^{eq} = -\frac{1}{\tau_g} g_\alpha^{(1)}, \tag{A2}$$

$$\kappa^2: \partial_{t_2} g_\alpha^{eq} + \left( \partial_{t_1} + \boldsymbol{e}_\alpha \cdot \nabla_1 \right) g_\alpha^{(1)} + \frac{\delta_t}{2} \left( \partial_{t_1} + \boldsymbol{e}_\alpha \cdot \nabla_1 \right)^2 g_\alpha^{eq} = -\frac{1}{\tau_g} g_\alpha^{(2)}. \tag{A3}$$

With Eq. (A2), Eq. (A3) can be rewritten as

$$\partial_{t_2} g_\alpha^{eq} + \left( 1 - \frac{\delta_t}{2\tau_g} \right) \left( \partial_{t_1} + \boldsymbol{e}_\alpha \cdot \nabla_1 \right) g_\alpha^{(1)} = -\frac{1}{\tau_g} g_\alpha^{(2)}. \tag{A4}$$

The zeroth- through third-order velocity moments of $g_\alpha^{eq}$ give the following equations:

$$\sum_\alpha g_\alpha^{eq} = (\varepsilon + P) \gamma^2 - P, \quad \sum_\alpha e_{\alpha i} g_\alpha^{eq} = (\varepsilon + P) \gamma^2 u_i, \tag{A5}$$

$$\sum_\alpha e_{\alpha i} e_{\alpha j} g_\alpha^{eq} = (\varepsilon + P) \gamma^2 u_i u_j + P \delta_{ij}, \tag{A6}$$

$$\sum_\alpha e_{\alpha i} e_{\alpha j} e_{\alpha k} g_\alpha^{eq} = (\varepsilon + P) \gamma^2 \left( u_k \delta_{ij} + u_i \delta_{jk} + u_j \delta_{ik} \right). \tag{A7}$$

According to the relationships $\sum_\alpha g_\alpha = \sum_\alpha g_\alpha^{eq}$ and $\sum_\alpha \boldsymbol{e}_\alpha g_\alpha = \sum_\alpha \boldsymbol{e}_\alpha g_\alpha^{eq}$, we have

$$\sum_\alpha g_\alpha^{(n)} = 0, \quad \sum_\alpha \boldsymbol{e}_\alpha g_\alpha^{(n)} = 0, \quad n = 1, 2, \cdots. \tag{A8}$$

Taking the summations of Eq. (A2) and Eq. (A4), we can obtain, respectively

$$\partial_{t_1}\left((\varepsilon+P)\gamma^2 - P\right) + \partial_{1i}\left((\varepsilon+P)\gamma^2 u_i\right) = 0, \tag{A9}$$

$$\partial_{t_2}\left((\varepsilon+P)\gamma^2 - P\right) = 0. \tag{A10}$$

Combining Eq. (A9) with Eq. (A10) leads to

$$\partial_t\left((\varepsilon+P)\gamma^2 - P\right) + \partial_i\left((\varepsilon+P)\gamma^2 u_i\right) = 0. \tag{A11}$$

Similarly, taking the first-order moments of Eq. (A2) and Eq. (A4), we can obtain, respectively

$$\partial_{t_1}\left((\varepsilon+P)\gamma^2 u_j\right) + \partial_{1j} P + \partial_{1i}\left((\varepsilon+P)\gamma^2 u_i u_j\right) = 0, \tag{A12}$$

$$\partial_{t_2}\left((\varepsilon+P)\gamma^2 u_j\right) + \left(1 - \frac{\delta_t}{2\tau_g}\right)\partial_{1i}\left(\sum_\alpha e_{\alpha i} e_{\alpha j} g_\alpha^{(1)}\right) = 0. \tag{A13}$$

According to Eq. (A2), $\sum_\alpha e_{\alpha i} e_{\alpha j} g_\alpha^{(1)}$ is given by

$$\sum_\alpha e_{\alpha i} e_{\alpha j} g_\alpha^{(1)} = -\tau_g\left[\partial_{t_1}\left(\sum_\alpha e_{\alpha i} e_{\alpha j} g_\alpha^{eq}\right) + \partial_{1k}\left(\sum_\alpha e_{\alpha i} e_{\alpha j} e_{\alpha k} g_\alpha^{eq}\right)\right]. \tag{A14}$$

With the aids of Eq. (A6) and Eq. (A7), we can obtain

$$\partial_{t_1}\left(\sum_\alpha e_{\alpha i} e_{\alpha j} g_\alpha^{eq}\right) = \partial_{t_1}\left(\sigma u_i u_j\right) + \partial_{t_1} P \delta_{ij}, \tag{A15}$$

$$\partial_{1k}\left(\sum_\alpha e_{\alpha i} e_{\alpha j} e_{\alpha k} g_\alpha^{eq}\right) = c_s^2 \partial_{1k}(\sigma u_k)\delta_{ij} + c_s^2 \sigma\left(\partial_{1i} u_j + \partial_{1j} u_i\right) + c_s^2\left(u_j \partial_{1i}\sigma + u_i \partial_{1j}\sigma\right), \tag{A16}$$

where $\sigma = (\varepsilon+P)\gamma^2$. According to Eq. (A12), $\partial_{t_1}(\sigma u_i u_j)$ is given by

$$\partial_{t_1}(\sigma u_i u_j) = u_j \partial_{t_1}(\sigma u_i) + u_i \partial_{t_1}(\sigma u_j) - u_i u_j \partial_{t_1}\sigma$$
$$= u_j\left[-\partial_{1k}(\sigma u_i u_k) - \partial_{1i} P\right] + u_i\left[-\partial_{1k}(\sigma u_j u_k) - \partial_{1j} P\right] - u_i u_j \partial_{t_1}\sigma, \tag{A17}$$

Neglecting the terms of $u^3$, Eq. (A17) can be written as

$$\partial_{t_1}(\sigma u_i u_j) = -u_j \partial_{1i} P - u_i \partial_{1j} P + O(u^3), \tag{A18}$$

The expression of $\partial_{t_1} P$ can be derived from Eq. (A9), and is given by

$$\partial_{t_1} P = -\frac{\partial_{1k}(\sigma u_k)}{(\varepsilon/P + 1)\gamma^2 - 1}. \tag{A19}$$

In the above derivation, $\partial_{t_1}\left((\varepsilon/P+1)\gamma^2\right)$ has been neglected. Substituting Eq. (A15) and Eq. (A16) together with Eq. (A18) and Eq. (A19) into Eq. (A14) gives

$$\sum_\alpha e_{\alpha i} e_{\alpha j} g_\alpha^{(1)} = -\tau_g \left\{ c_s^2 \sigma \left(\partial_{1i} u_j + \partial_{1j} u_i\right) + \chi c_s^2 \partial_{1k}\left(\sigma u_k\right)\delta_{ij} + \left[u_j \partial_{1i}\left(c_s^2\sigma - P\right) + u_i \partial_{1j}\left(c_s^2\sigma - P\right)\right]\right\}, \quad (A20)$$

where $\chi = 1 - P/\left[c_s^2(\sigma - P)\right]$. Substituting Eq. (A20) into Eq. (A13) and then combining Eq. (A13) with Eq. (A12), we can obtain

$$\partial_t\left((\varepsilon+P)\gamma^2 u_j\right) + \partial_i\left((\varepsilon+P)\gamma^2 u_i u_j\right) = -\partial_j P + \partial_i \Pi_{ij} + O(u^3), \quad (A21)$$

where $\Pi_{ij}$ is the stress tensor and is given by

$$\Pi_{ij} = \eta\left[\gamma\left(\partial_i u_j + \partial_j u_i\right) - \frac{2}{D}\gamma \partial_k u_k \delta_{ij}\right] + \varsigma \gamma \partial_k u_k \delta_{ij} + (\tau_g - 0.5\delta_t)\chi c_s^2 \partial_k(\sigma u_k)\delta_{ij}$$
$$+ (\tau_g - 0.5\delta_t)\left[u_j \partial_i\left(c_s^2\sigma - P\right) + u_i \partial_j\left(c_s^2\sigma - P\right)\right], \quad (A22)$$

where $\eta = (\tau_g - 0.5\delta_t) c_s^2 (\varepsilon + P)\gamma$ is the shear viscosity and $\varsigma = 2\eta/D$ is the bulk viscosity. The first and second terms on the right-hand side of Eq. (A22) can be rewritten as follows:

$$\eta\left[\gamma\left(\partial_i u_j + \partial_j u_i\right) - \frac{2}{D}\gamma \partial_k u_k \delta_{ij}\right] + \varsigma \gamma \partial_k u_k \delta_{ij} = \eta\left[\partial_i(\gamma u_j) + \partial_j(\gamma u_i) - \frac{2}{D}\partial_k(\gamma u_k)\delta_{ij}\right] + \varsigma \partial_k(\gamma u_k)\delta_{ij}$$
$$- \eta\left[u_i \partial_j \gamma + u_j \partial_i \gamma - \frac{2}{D} u_k \partial_k \gamma \delta_{ij}\right] - \varsigma u_k \partial_k \gamma \delta_{ij}. \quad (A23)$$

Since the derivative of the Lorentz factor $\partial_j \gamma$ is proportional to $u_i \partial_j u_i$, the last two terms on the right-hand side of Eq. (A23) will be of the order $u^3$. Then Eq. (A22) can be rewritten as

$$\Pi_{ij} = \eta\left[\partial_i(\gamma u_j) + \partial_j(\gamma u_i) - \frac{2}{D}\partial_k(\gamma u_k)\delta_{ij}\right] + \varsigma \partial_k(\gamma u_k)\delta_{ij} + (\tau_g - 0.5\delta_t)\chi c_s^2 \partial_k(\sigma u_k)\delta_{ij}$$
$$+ (\tau_g - 0.5\delta_t)\left[u_j \partial_i\left(c_s^2\sigma - P\right) + u_i \partial_j\left(c_s^2\sigma - P\right)\right] + O(u^3), \quad (A24)$$

Finally, we give the expression of $\partial_{t_1} u_j$, which is needed in the Chapman-Enskog analysis of Eq. (5). From Eq. (A12), we have

$$(\varepsilon+P)\gamma^2 \partial_{t_1} u_j + u_j \partial_{t_1}\left((\varepsilon+P)\gamma^2\right) + \partial_{1j} P + u_j \partial_{1i}\left((\varepsilon+P)\gamma^2 u_i\right) + (\varepsilon+P)\gamma^2 u_i \partial_{1i} u_j = 0. \quad (A25)$$

Then the following equation can be obtained by substituting Eq. (A9) into Eq. (A25):

$$(\varepsilon+P)\gamma^2 \partial_{t_1} u_j - u_j \partial_{t_1} P + \partial_{1j} P + (\varepsilon+P)\gamma^2 u_i \partial_{1i} u_j = 0. \quad (A26)$$

Hence $\partial_{t_1} u_j$ is given by

$$\partial_{t_1} u_j = -u_i \partial_{1i} u_j - \frac{\left(-u_j \partial_{t_1} P + \partial_{1j} P\right)}{(\varepsilon + P)\gamma^2}, \quad (A27)$$

According to Eq. (A19), for weakly relativistic problems ($1 < \gamma \ll 2$ and $|u/c_l| < 0.3$), $u_j \partial_{t_1} P$ will be much smaller than $\partial_{1j} P$ and can be neglected.

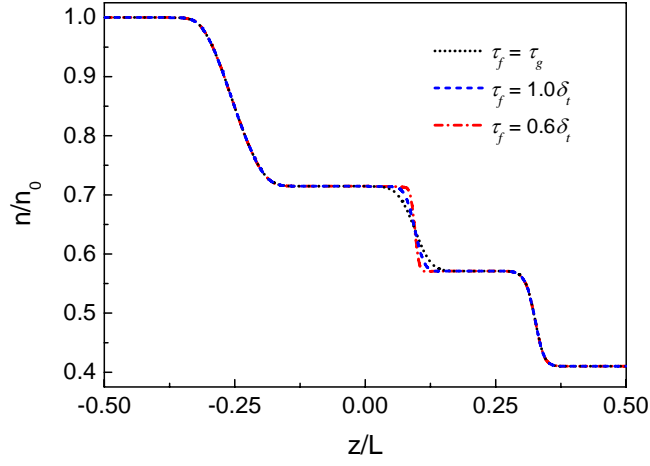

(a) $\eta/s = 0.01$

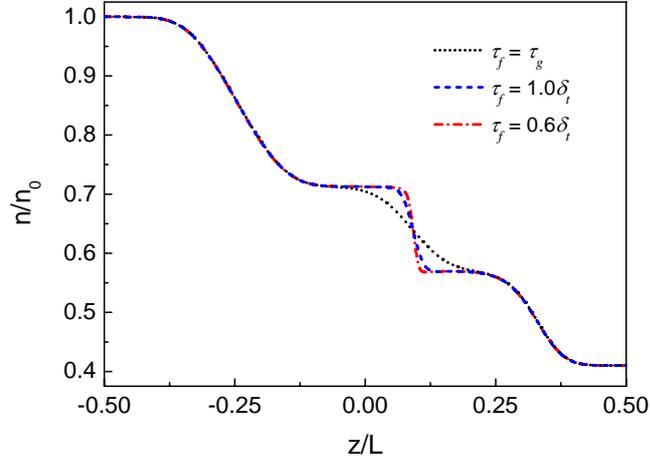

(b) $\eta/s = 0.05$

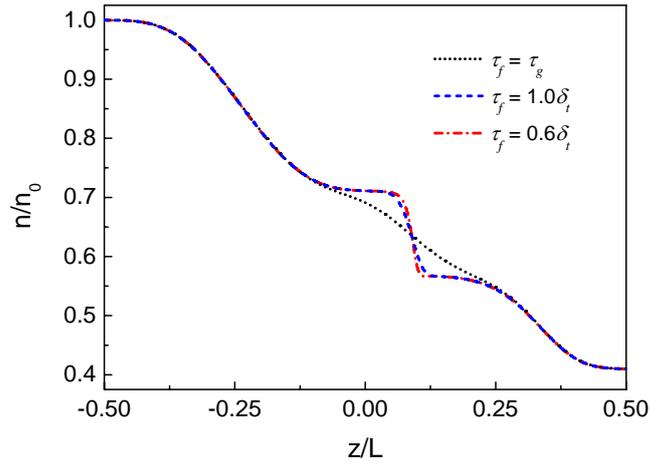

(c) $\eta/s = 0.1$

FIG. 1. Particle numer profiles at $t = 400\delta_t$ with different values of $\tau_f$: (a) $\eta/s = 0.01$, (b) $\eta/s = 0.05$, and (c) $\eta/s = 0.1$.

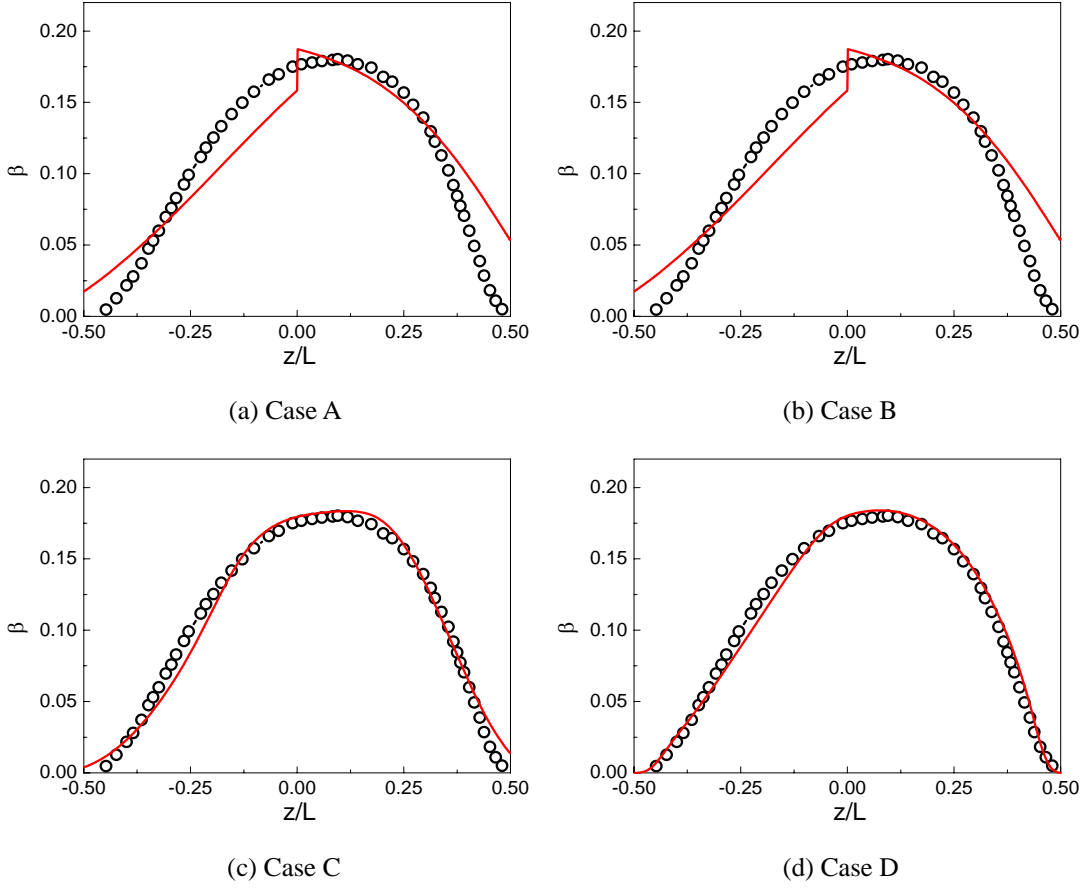

FIG. 2. Velocity profiles of weakly relativistic shock wave in quark-gluon plasma at $\eta/s = 0.5$ in different cases. Case A: $\tau_e = \tau_\varepsilon = \tau_q = \tau_v$; Case B: $\tau_e = \tau_\varepsilon = \tau_v$, $\tau_q - 0.5\delta_t = a(\tau_v - 0.5\delta_t)$; Case C: $\tau_\varepsilon = \tau_q = \tau_v$, $\tau_e - 0.5\delta_t = a(\tau_v - 0.5\delta_t)$; and Case D: $\tau_q = \tau_v$, $\tau_e = \tau_\varepsilon = 0.5\delta_t + a(\tau_v - 0.5\delta_t)$. The circles represent the results of BAMPS.

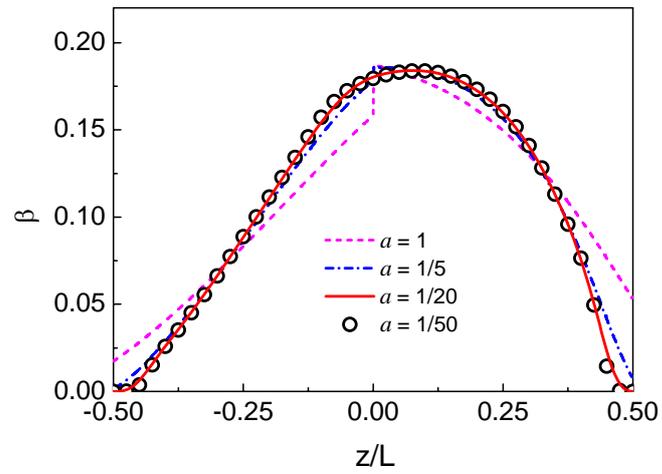

FIG. 3. Velocity profiles of Case D at $\eta/s = 0.5$ with different values of $a$.

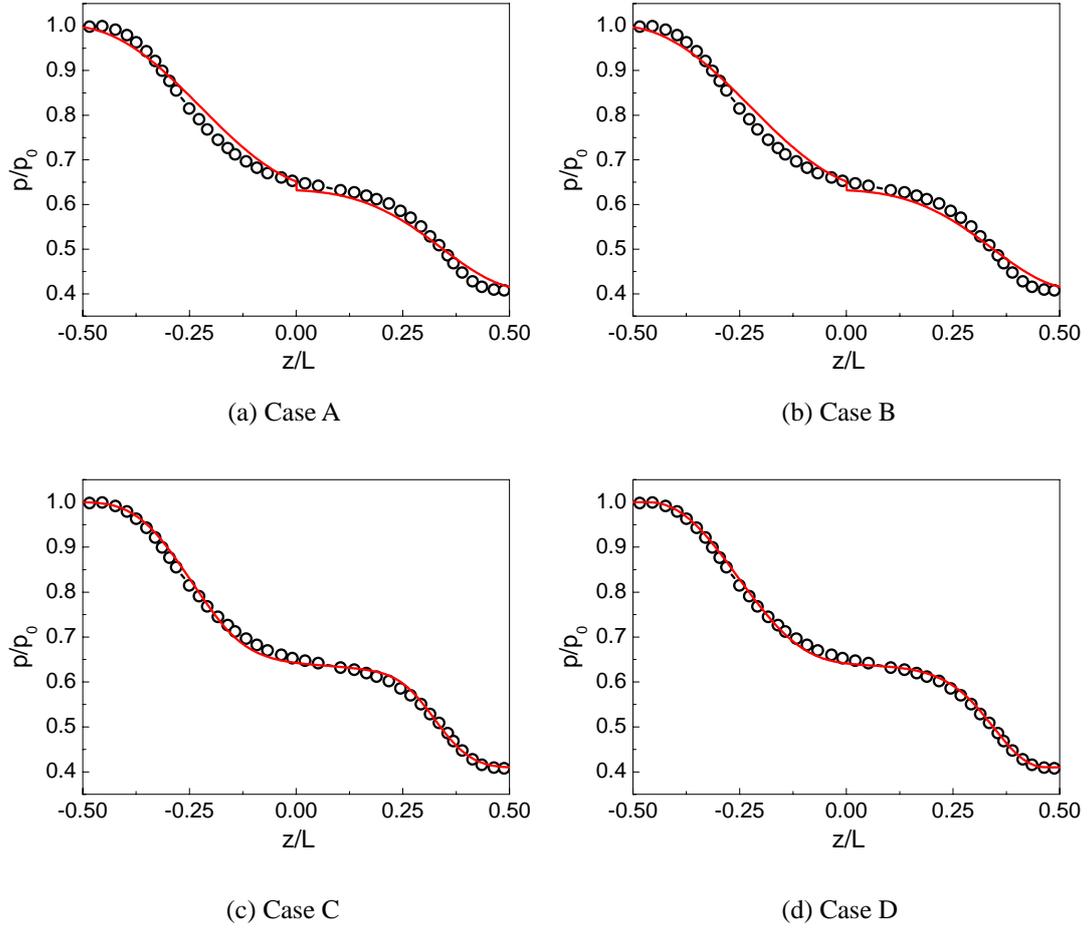

(a) Case A

(b) Case B

(c) Case C

(d) Case D

FIG. 4. Pressure profiles of weakly relativistic shock wave in quark-gluon plasma at $\eta/s = 0.2$ in different cases.

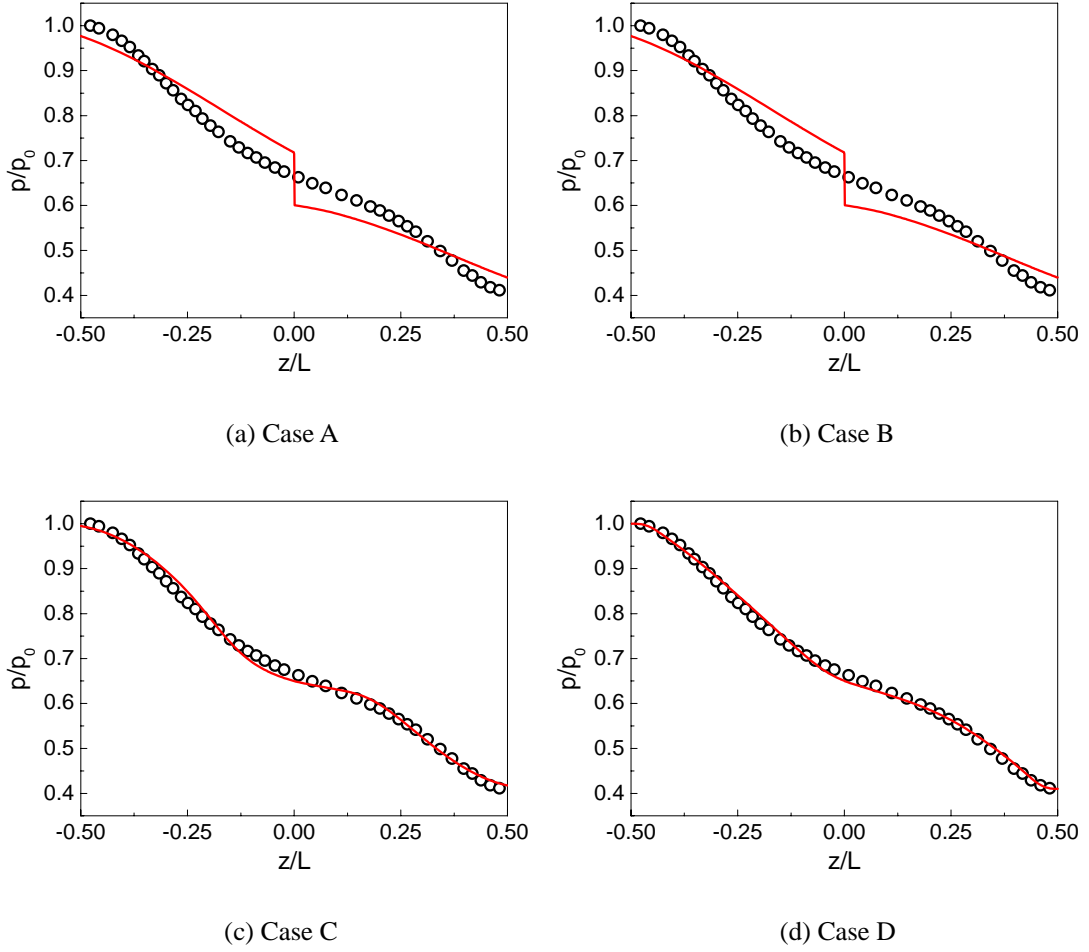

(a) Case A

(b) Case B

(c) Case C

(d) Case D

FIG. 5. Pressure profiles of weakly relativistic shock wave in quark-gluon plasma at $\eta/s = 0.5$ in different cases.

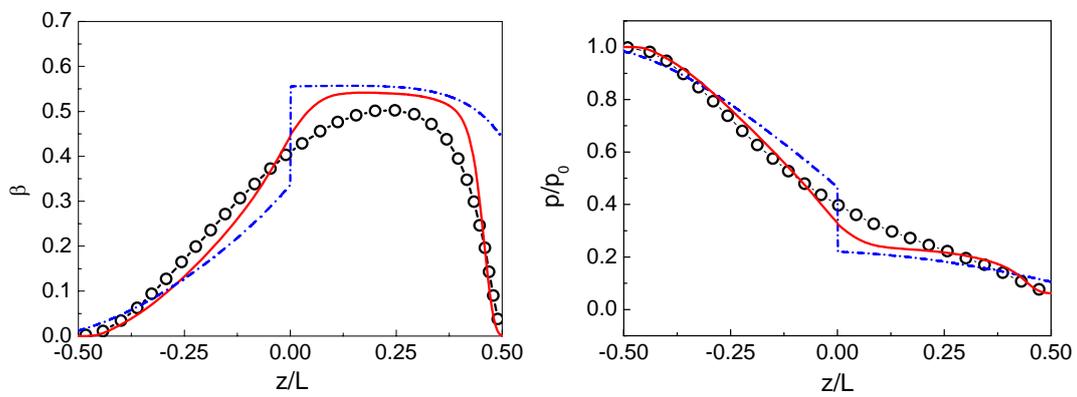

FIG. 6. Velocity (left) and pressure (right) profiles of moderately relativistic shock wave in quark-gluon plasma at $\eta/s = 0.5$. The dash-dottied lines and the solid lines represent the results of Cases A and D, respecitvely. The circles represent the results of BAMPS.

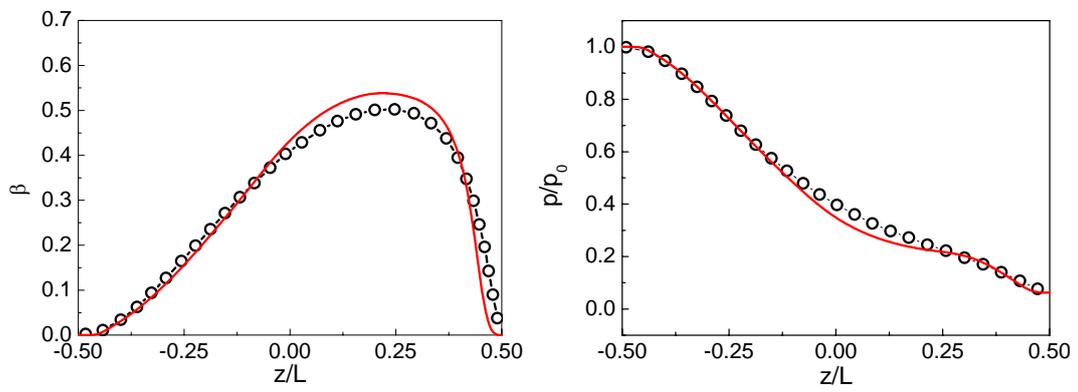

FIG. 7. Same as FIG. 6. Corrected velocity (left) and pressure (right) profiles obtained with Eqs. (37) and (38).